\documentclass[3p,times,twocolumn]{elsarticle}

\usepackage{ecrc}

\volume{00}

\firstpage{1}

\journalname{Nuclear and Particle Physics Proceedings}

\runauth{}

\jid{nppp}

\jnltitlelogo{Nuclear and Particle Physics Proceedings}

\usepackage{amssymb}
\usepackage{lineno}
\usepackage[version=4]{mhchem}
\usepackage{hyperref}
\usepackage{mwe}
\usepackage{subfig}
\usepackage{booktabs}
\usepackage{tabularx}
\usepackage{graphicx}
\graphicspath{{images/}}
\usepackage[figuresright]{rotating}
\usepackage[skip=0pt]{caption} 

\begin{document}

\begin{frontmatter}

\dochead{}

\title{Eco-friendly Resistive Plate Chambers for detectors in future HEP applications}
\vspace{0pt}
\author[j]{Luca Quaglia* on behalf of the RPC EcoGas@GIF++ collaboration}
\author[a]{R. Cardarelli, B. Liberti, E. Pastori, G. Proto, L. Pizzimento, A. Rocchi}
\author[b]{G. Aielli, P. Camarri, A. Di Ciacco, L. Di Stante, R. Santonico}
\author[c]{D. Boscherini, A. Bruni, L. Massa, A. Polini, M. Romano}
\author[d]{L. Benussi, S. Bianco, D. Piccolo}
\author[e]{G. Saviano}
\author[f]{M. Abbrescia, L. Congedo, M. De Serio, G. Galati, G. Pugliese, S. Simone, D. Ramos}
\author[g]{P. Salvini}
\author[h]{A. Samalan, M. Tytgat, N. Zaganidis} 
\author[i]{J. Eysermans}
\author[j]{A. Ferretti, M. Gagliardi, L. Terlizzi, E. Vercellin}
\author[k]{P. Dupieux, B. Joly, S.P. Manen}
\author[l]{R. Guida, B. Mandelli, G. Rigoletti}
\author[m]{M. Barroso}
\author[n]{M.C. Arena}
\author[o]{A. Pastore}
\author[o,p]{R. Aly}
\author[q]{M. Verzeroli}
\author[r]{S. Buontempo}

\address[a]{INFN sezione di Roma Tor Vergata, Rome, Italy}

\address[b]{Dipartimento di Fisica di Roma Tor Vergata, Rome, Italy}

\address[c]{INFN sezione di Bologna, Bologna, Italy}

\address[d]{Laboratori Nazionali di Frascati dell'INFN, Frascati, Italy}

\address[e]{Dipartimento di Ingegneria Chimica Materiali Ambiente, Sapienza Università di Roma, Rome, Italy}

\address[f]{Dipartimento di Fisica di Bari e sezione INFN di Bari, Bari, Italy}

\address[g]{Sezione INFN di Pavia, Pavia, Italy}

\address[h]{Dept. of Physics and Astronomy, Ghent University, Ghent, Belgium } 

\address[i]{MIT, Cambridge, Massachussets, U.S.A.}

\address[j]{Università degli studi di Torino and INFN, sezione di Torino, Torino, Italy}

\address[k]{Clermont Univeristé, Université Blaise Pascal, CNRS/IN2P3, Laboratoire de Physique Corpusculaire, Clermont-Ferrand, France}

\address[l]{CERN, Geneve, Switzerland}

\address[m]{Universidade do Estdo do Rio de Janeiro, Rio de Janeiro, Brazil}

\address[n]{Università degli Studi di Pavia, Pavia, Italy}

\address[o]{Sezione INFN di Bari, Bari, Italy}

\address[p]{Helwan University, Ain Helwan 11795 Cairo, Egypt}

\address[q]{Université Clause Bernard Lyon I, France}

\address[r]{Sezione INFN di Napoli, Napoli, Italy}

\address{*Via P. Giuria 1 Turin, Italy}
\vspace{-10pt}
\begin{abstract}
Resistive Plate Chamber detectors are largely used in current High Energy Physics experiments, typically operated in avalanche mode with large fractions of Tetrafluoroethane (C2H2F4), a gas recently banned by the European Union due to its high Global Warming Potential (GWP).
An intense R\&D activity is ongoing to improve RPC technology in view of future HEP applications.
In the last few years the RPC EcoGas@GIF++ Collaboration has been putting in place a joint effort between the ALICE, ATLAS, CMS, LHCb/SHiP and EP-DT Communities to investigate the performance of present and future RPC generations with eco-friendly gas mixtures.
Detectors with different layout and electronics have been operated with ecological gas mixtures, with and without irradiation at the CERN Gamma Irradiation Facility (GIF++). Results of these performance studies together with plans for an aging test campaign are discussed in this article.
\end{abstract}


\end{frontmatter}


\section{Introduction}
\label{sec.intro}

Resistive Plate Chambers (RPCs) are gaseous detectors widely employed in the muon systems of high energy physics experiments. Thanks to their timing and spatial resolution of 1-2 ns and few mm respectively, they can be employed for muon triggering and identification purposes. Furthermore, due to their comparatively low cost, they are suitable to cover large areas. This is the case for the RPCs in the muon systems of the ATLAS, ALICE and CMS experiments.

These detectors are operated with gas mixtures that contain different percentages of \ce{C_{2}H_{2}F_{4}} (R134a), \ce{SF_{6}} and \ce{i-C_{4}H_{10}}. Of these three gases, \ce{C_{2}H_{2}F_{4}} and \ce{SF_{6}} are greenhouse gases (F-gases) with high Global Warming Potential (GWP), meaning that they trap great amounts of heat in the atmosphere for an extended period of time. European Union regulations\cite{eu-517-2014} have imposed a progressive phase down in the production and use of such gases, reducing their availability on the market and increasing their price. Moreover, RPCs are the main contributor to the emissions of F-gases at CERN, due to leaks at the detector level. For these reasons, the search of a more eco-friendly gas mixture is of the greatest importance.

R134a is the greatest contributor to the mixture GWP, so the first step is to find a substitute for this gas. A promising candidate could be \textit{tetrafluoropropene} (\ce{C_{3}H_{2}F_{4}} or HFO-1234ze or simply HFO). In LHC systems, this gas cannot fully replace R134a, since the detector working point would move to over 15 kV. For this reason, HFO has to be diluted with a \textit{placeholder} gas, that does not directly take part in the ionization process but increases the mean free path, effectively lowering the working point. Following many laboratory studies with cosmic rays\cite{prelPiccolo,perlLiberti,prelGuida,prelAntonio,prelShip}, the \textit{RPC EcoGas@GIF++} collaboration was born. It is composed by members of the four LHC experiments and the CERN Detector Technology group and it is focused on the full replacement of R134a with HFO, using \ce{CO_{2}} as placeholder gas. The focus of the collaboration is to perform long-term irradiation tests, to study the stability of the detectors operated with these new gas mixtures and to carry out performance studies in controlled environment, such as beam tests.

\section{Experimental setup}
\label{sec.setup}

\begin{figure} [h!]
	\centering
	\includegraphics[height = 0.6\linewidth]{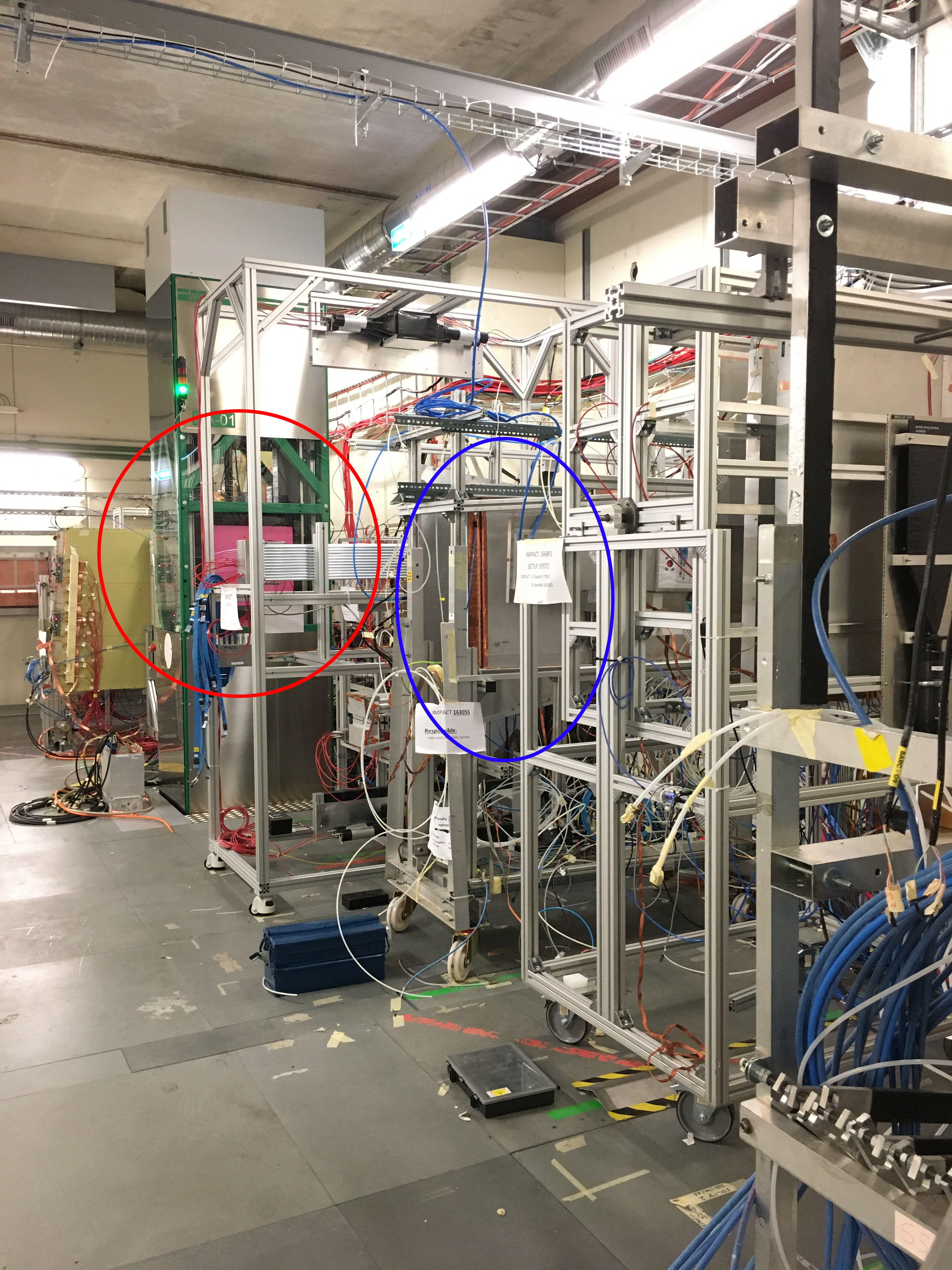}
	\caption{Picture of the EcoGas setup at GIF++. The irradiation source is in the red circle, while the RPCs are highlighted in blue}
	\label{fig.setup}
\end{figure}

Each of the groups has provided an RPC prototype, that was installed on one of two mechanical frames inside the bunker of the Gamma Irradiation Facility (GIF++) at CERN\cite{gifGeneral}. This facility is equipped with a 12.5 TBq \ce{^{137}Cs} source, which can be exploited to simulate a high radiation background on the detectors. The photon flux can be modulated by means of lead filters, in order to test the detectors in different irradiation conditions (referred to as ABS in the following). For reference, each ABS is represented by a pure number that roughly represents how much the radiation is attenuated, with respect to the no-shielding condition (i.e. ABS 1 corresponds to maximum radiation, while ABS 46000 to the minimum). Furthermore, in dedicated beam time periods, a high energy (100 GeV/c) muon beam is available in order to test the RPCs' performance. The peculiarity of GIF++ is the possibility to combine the muon beam with the \ce{^{137}Cs} source, to test also the rate capability of the detectors. In figure \ref{fig.setup}, a photo of the experimental setup is shown. 

The two mechanical structures on which the detectors are installed are located at 3 and 6 meters from the source, due to the different experimental conditions where the various detectors will be eventually operated. THE CMS, EP-DT and ATLAS RPCs are placed on \textit{trolley 3}, at 3 m from the source, while the ALICE and SHiP RPCs are on \textit{trolley 1}, at 6 m from the source. The characteristics of each detector are listed in table \ref{tab.detectors}.

\begin{table}[h!]
	\begin{center}
		\setlength{\tabcolsep}{1.5pt} 
		\begin{tabular}{|c|c|c|c|c|c|}
			\toprule
			\textbf{Detector} & \textbf{Area (cm$^{2}$)} & \textbf{Gaps} & \textbf{Thickness} & \textbf{Readout} & \textbf{Strips} \\
			\toprule
			ALICE & 2500 & 1 & 2 mm & TDC & 32\\
			
			ATLAS & 550 & 1 & 2/1.8 mm & Digitizer & 1\\
			
			CMS & 4350 & 2 & 2 mm & TDC & 128\\
			
			EP-DT & 7000 & 1 & 2 mm & Digitizer & 7\\
		
			SHiP & 7000 & 1 & 1.6 mm & TDC & 64\\
			\bottomrule
		\end{tabular}
		\caption{Features of the collaboration's detectors}
		\label{tab.detectors}
	\end{center}
\end{table}

\vspace{-20pt}
In order to have a detector-independent estimate of the irradiation rate of all RPCs, it was decided to carry out independent measurements of the instant dose rate (in $\mu$S/h) using a \href{https://www.mirion.com/products/rds-31-itx-telemetry-survey-meters}{Mirion RDS-31ITX} dosimeter. Thanks to this, we managed to understand that the following pairs of absorption factors correspond to the same instant dose at the two distances: ABS 10 and 69 (500 $\mu$Sv/h) and 2.2 and 22 (2000 $\mu$Sv/h) at 6 and 3 m respectively. For this reason, in the following, all the results shown will correspond to these values of ABS.

During the beam test, we tested two different eco-friendly gas mixtures, plus the one currently employed in ATLAS and CMS, referred to as the standard gas mixture. The two ecological mixtures are called ECO2 and ECO3 (since they are the second and third gas mixtures tested by the collaboration). The composition of all the tested gas mixtures is listed in table \ref{tab.conc}.

\begin{table}
	\begin{center}
		\setlength{\tabcolsep}{2pt} 
		\begin{tabular}{|c|c|c|c|c|c|}
			\toprule
			\textbf{Name} & \textbf{\ce{C_{2}H_{2}F_{4}}} & \textbf{\ce{CO_{2}}} & \textbf{HFO} & \textbf{\ce{i-C_{4}H_{10}}} & \textbf{\ce{SF_{6}}} \\
			\toprule
			STD & 95.2 & 0 & 0 & 4.5 & 0.3\\
			
			ECO2 & 0 & 60 & 35 & 4 & 1\\
			
			ECO3 & 0 & 69 & 25 & 5 & 1\\
			\bottomrule
		\end{tabular}
		\caption{Composition of the mixtures tested}
		\label{tab.conc}
	\end{center}
\end{table}

\vspace{-10pt}

\section{Beam test results}
\label{sec.beam}

The trigger for the beam test was provided by the coincidence of four scintillators, coupled with photomultipliers. Two of them were placed inside the GIF++ bunker and two outside, providing a trigger area of 10x10 cm$^{2}$. In the following, results obtained from different detectors will be shown.
 
\subsection{Charge spectra}
\label{sub.charge}

\begin{figure} [!h]

\begin{minipage}{.45\linewidth}
	\subfloat[STD]{\label{Standard}\includegraphics[width = 1\linewidth]{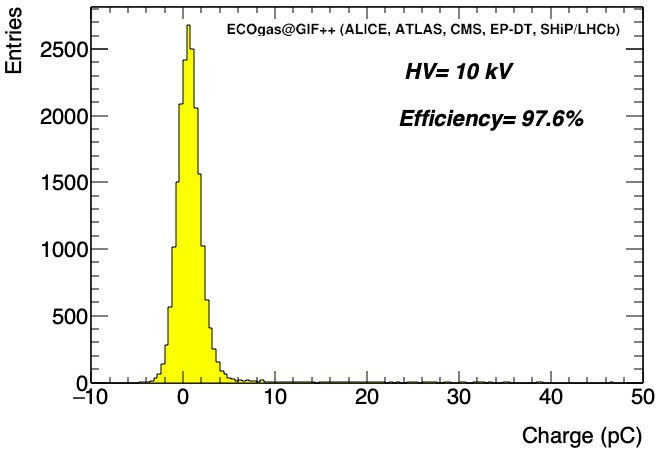}}
\end{minipage}%
\hspace{0.1pt}
\begin{minipage}{.45\linewidth}
	\centering
	\subfloat[ECO2]{\label{ECO2}\includegraphics[width = 1\linewidth]{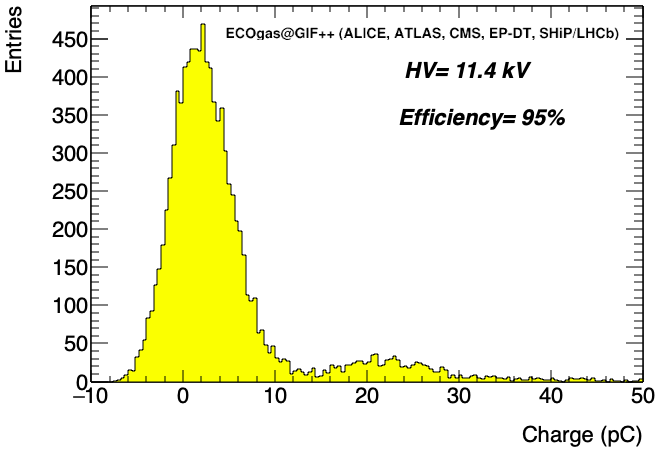}}
\hspace{0pt}
\end{minipage}
\centering
\begin{minipage}{.45\linewidth}
	
	\subfloat[ECO3]{\label{ECO3}\includegraphics[width = 1\linewidth]{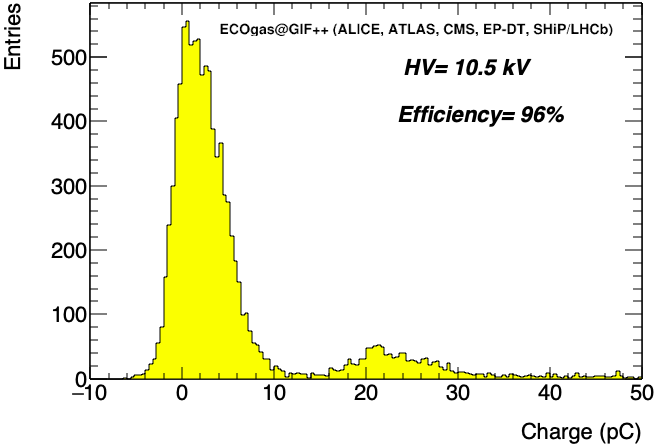}}
\end{minipage}%
\caption{Results from ATLAS chamber: charge distribution for the three mixtures tested}
\label{fig.charge}
\end{figure}

The ATLAS RPC (2 mm single gas gap and 1.8 mm electrode), is read out by a digitizer, allowing one to calculate the prompt charge of a signal. The results of this measurement are shown in figure \ref{fig.charge}.

\hspace{0pt}
\vspace{-10pt}

The spectra are shown for the same efficiency values and without irradiation (source off). The top left figure refers to the standard gas mixture, where a single and well defined peak is observed, without streamer contamination. For the ECO2 and ECO3, a secondary peak appears at higher values of charge due to the presence of secondary avalanches (transition signals). The increase of mean signal charge could lead to an acceleration of the aging process and, at the same time, to a reduction of the detectors' rate capability. The presence of events with a negative charge can be explained by considering that the integration window coincides with the digitizer acquisition window and the presence of a low frequency noise induced in the signal cables can lead to non-positive integrals of the RPC signals. 

\subsection{Efficiency, currents and cluster size}
\label{sub.eff}

\begin{figure} [!h]
	
	\begin{minipage}{\linewidth}
		\centering
		\subfloat[STD]{\label{EffStandard1_6}\includegraphics[height = 0.4\linewidth]{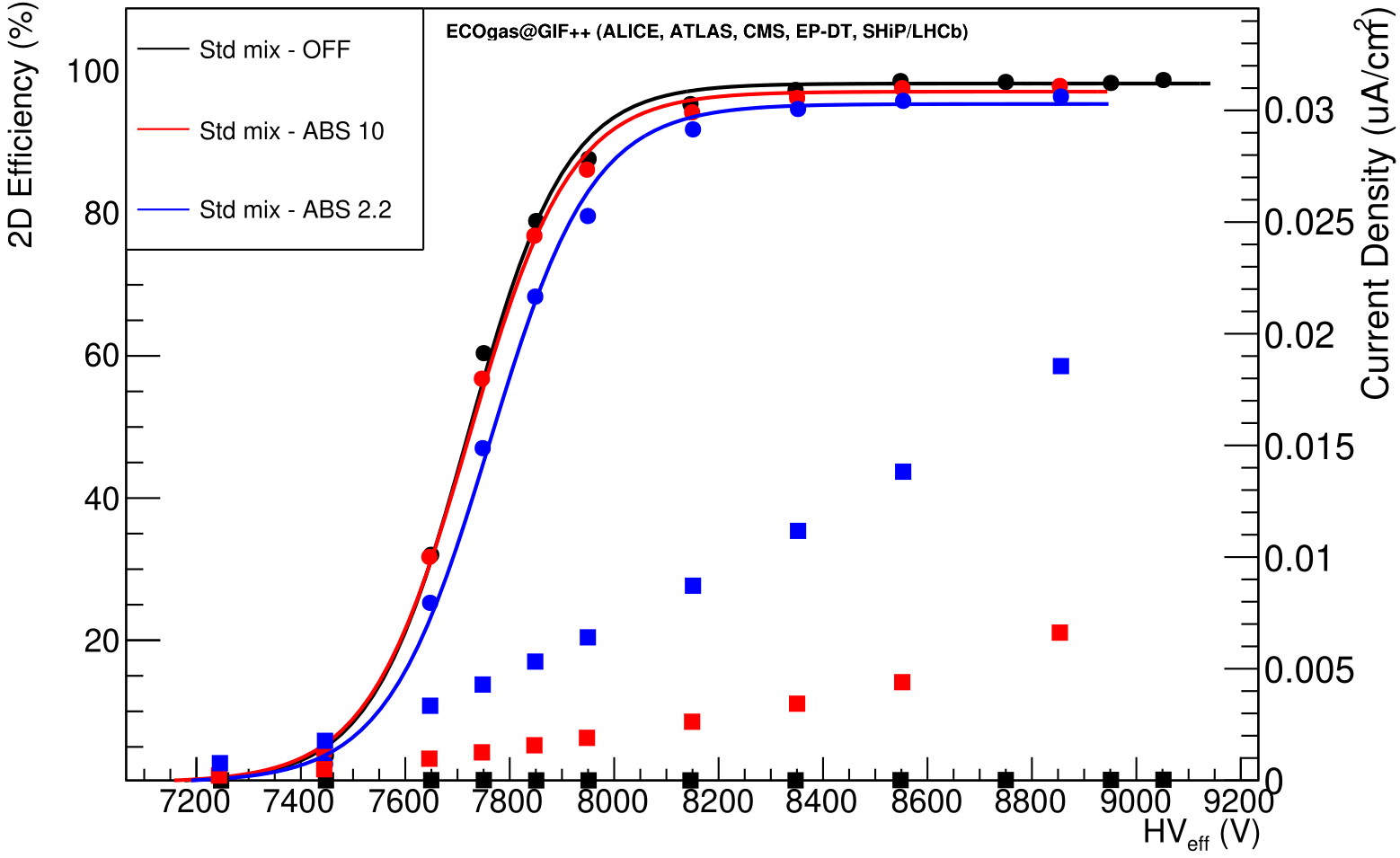}}
	\end{minipage}%
	\hspace{0.1pt}
	\begin{minipage}{\linewidth}
		\centering
		\subfloat[ECO2]{\label{EffECO21_6}\includegraphics[height = 0.4\linewidth]{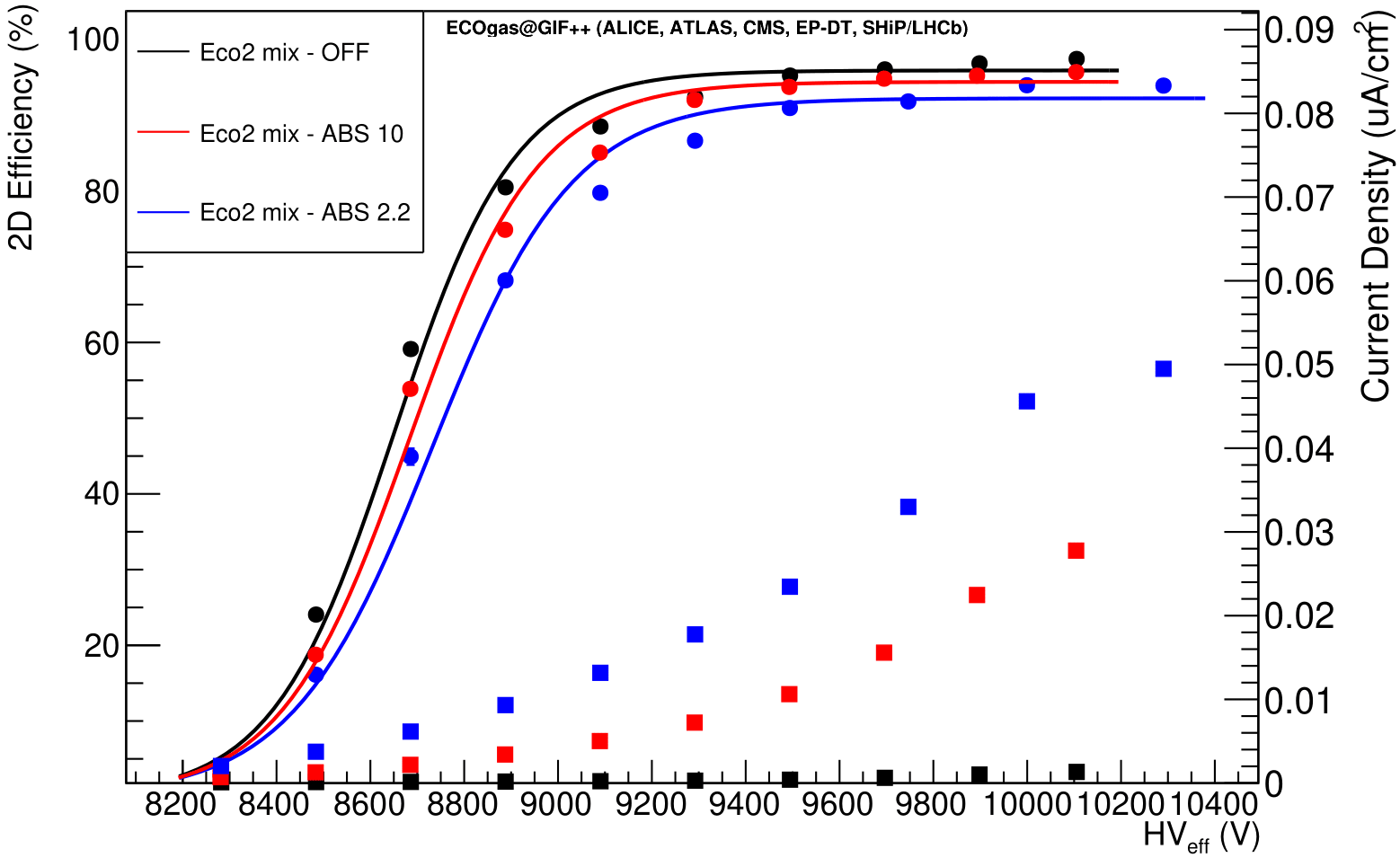}}
		\hspace{0pt}
	\end{minipage}
	\hspace{0pt}
	\centering
	\begin{minipage}{\linewidth}
		\centering
		\subfloat[ECO3]{\label{EffECO31_6}\includegraphics[height = 0.4\linewidth]{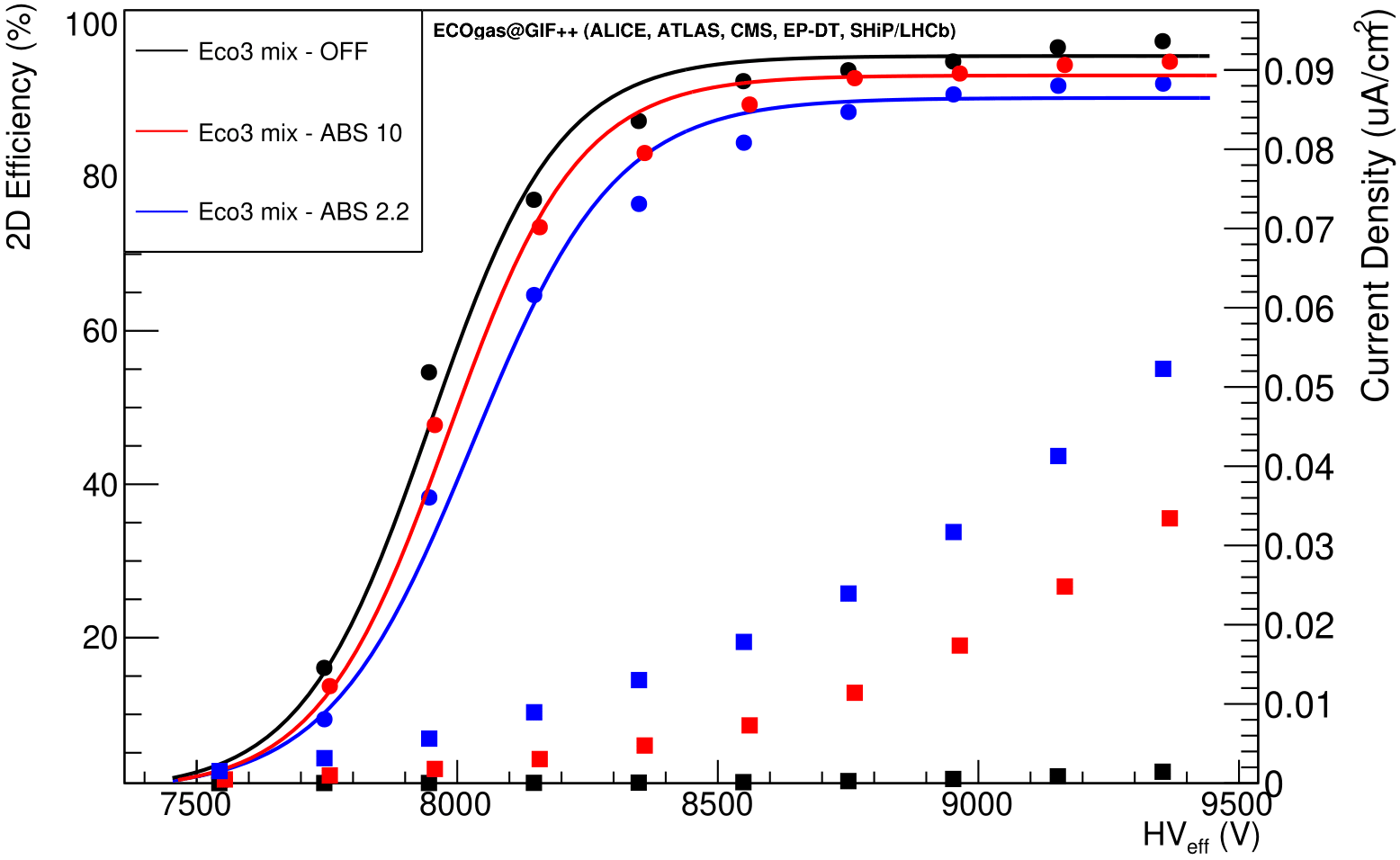}}
	\end{minipage}%
	\caption{Results from SHiP chamber: efficiency and current density as a function of the high voltage, for the three mixtures tested}
	\label{fig.eff}
\end{figure}

The results of efficiency measurements will be presented for two RPCs, the SHiP/LHCb and EP-DT ones. They are both single-gap detectors with different gap and electrodes thickness: the former has is 1.6 mm thick gap, while the latter is 2 mm. Since the SHiP RPC is equipped with two strip planes, it is considered efficient if, for a given trigger, at least a hit is detected in both strip planes. The SHiP RPC is readout using the new front-end electronics that have been developed for the ALICE experiment, the FEERIC FEE\cite{FEERIC}. The results of these measurements are shown in figure \ref{fig.eff}, together with the trends of absorbed current.

\hspace{0pt}

\begin{figure} [!h]
	\begin{minipage}{\linewidth}
		\centering
		\subfloat[STD]{\label{EffStandard2mm}\includegraphics[height = 0.6\linewidth]{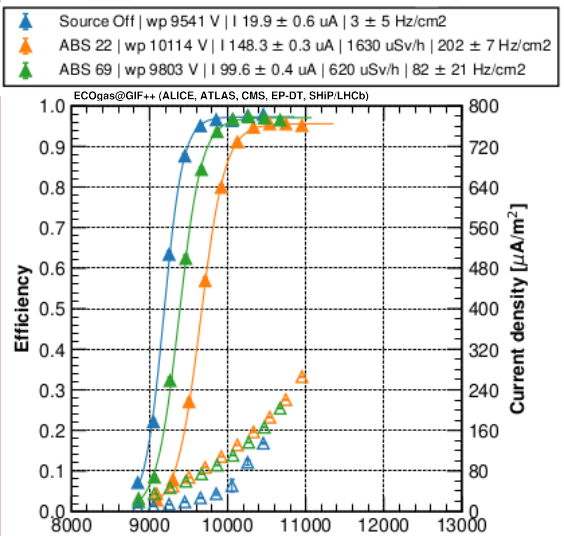}}
	\end{minipage}%
	\hspace{0.1pt}
	\begin{minipage}{\linewidth}
		\centering
		\subfloat[ECO2]{\label{EffECO22mm}\includegraphics[height = 0.6\linewidth]{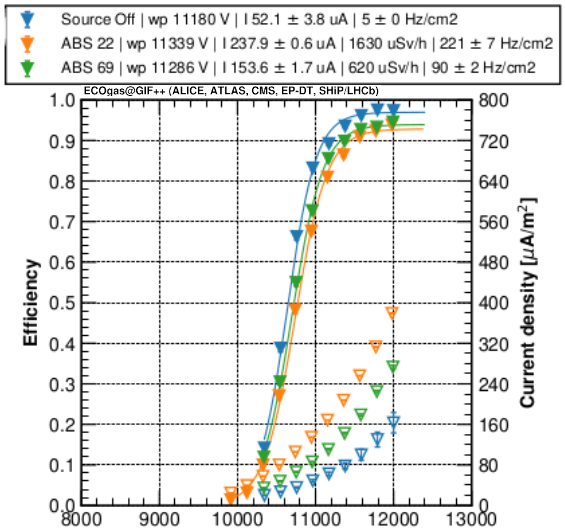}}
		\hspace{0pt}
	\end{minipage}
	\hspace{0pt}
	\centering
	\begin{minipage}{\linewidth}
		\centering
		\subfloat[ECO3]{\label{EffECO32mm}\includegraphics[height = 0.6\linewidth]{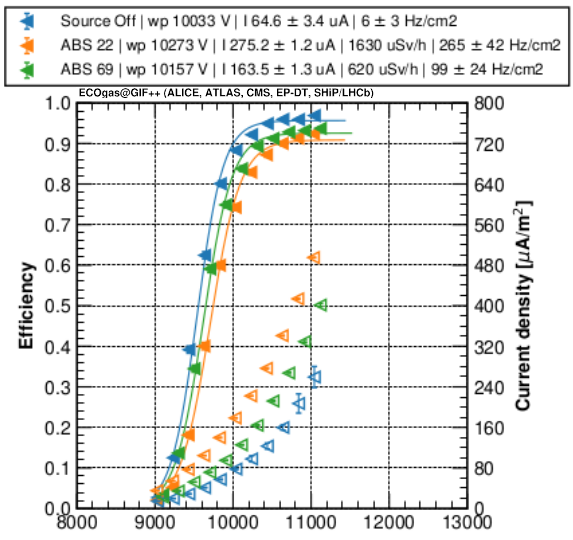}}
	\end{minipage}%
	\caption{Results from EP-DT chamber: efficiency and current density as a function of the high voltage, for the three mixtures tested}
	\label{fig.effEpDt}
\end{figure}

The three figures refer to the different gas mixtures that have been tested and to three irradiation conditions that have been selected to compare the performance on the two trolleys. The working point, for a given mixture, tends to shift towards higher values for higher irradiation and also the maximum value of efficiency reached gets lower. Both of these effects can be attributed to the rate capability of the detector. Going from the standard gas mixture to the eco-friendly ones, one can observe a shift of the source off curve of around 1 and 0.4 kV, for ECO2 and ECO3 respectively. Although the maximum efficiency reached at source off is comparable among all the mixtures, the drop with maximum irradiation is much greater for ECO2 and ECO3, 4 and 6\% respectively. Lastly, the absorbed current is doubled with respect to the standard gas mixture. This effect has to be closely monitored and could be linked with the increase of mean prompt charge released in the gas. Very similar observations can be made by looking at the results for the EP-DT RPC, with 2 mm gas gap. The same curves are shown in figure \ref{fig.effEpDt}.

\hspace{0pt}
\vspace{0pt}
The efficiency drop is very similar to the 1.6 mm case, while the current absorbed without irradiation is double with respect to the standard gas mixture with ECO2 and tripled with ECO3. Under irradiation the increase of current is 1.5 times for ECO2 and 1.8 times for ECO3.

Another important quantity that has been measured in beam tests is the muon cluster size, i.e. the number of adjacent strips that are fired per particle signal. The results are quite similar for all the detectors under test and the trend is very similar for all gas mixtures. In particular, a slight decrease of the average cluster size, with increasing irradiation, is observed. This can be explained considering that the real voltage between the gaps is effectively reduced by a factor R$\cdot$I where R is the resistance of the electrodes and I is the current circulating in the detectors, so the avalanche development is reduced and, in terms, the cluster size is smaller. Lastly, the average cluster size does not seem to depend on the gas mixture and, at the working point, the value is compatible across all the mixtures.   

\subsection{Preliminary aging campaign}
\label{sub.aging}

In order to monitor the detector operation on the long-term an aging test is being carried out. During these studies, the detectors are switched on to a fixed value of high voltage and they are exposed to the photon flux from the gamma source.  

\begin{figure} [h!]
	\centering
	\includegraphics[height = 0.5\linewidth]{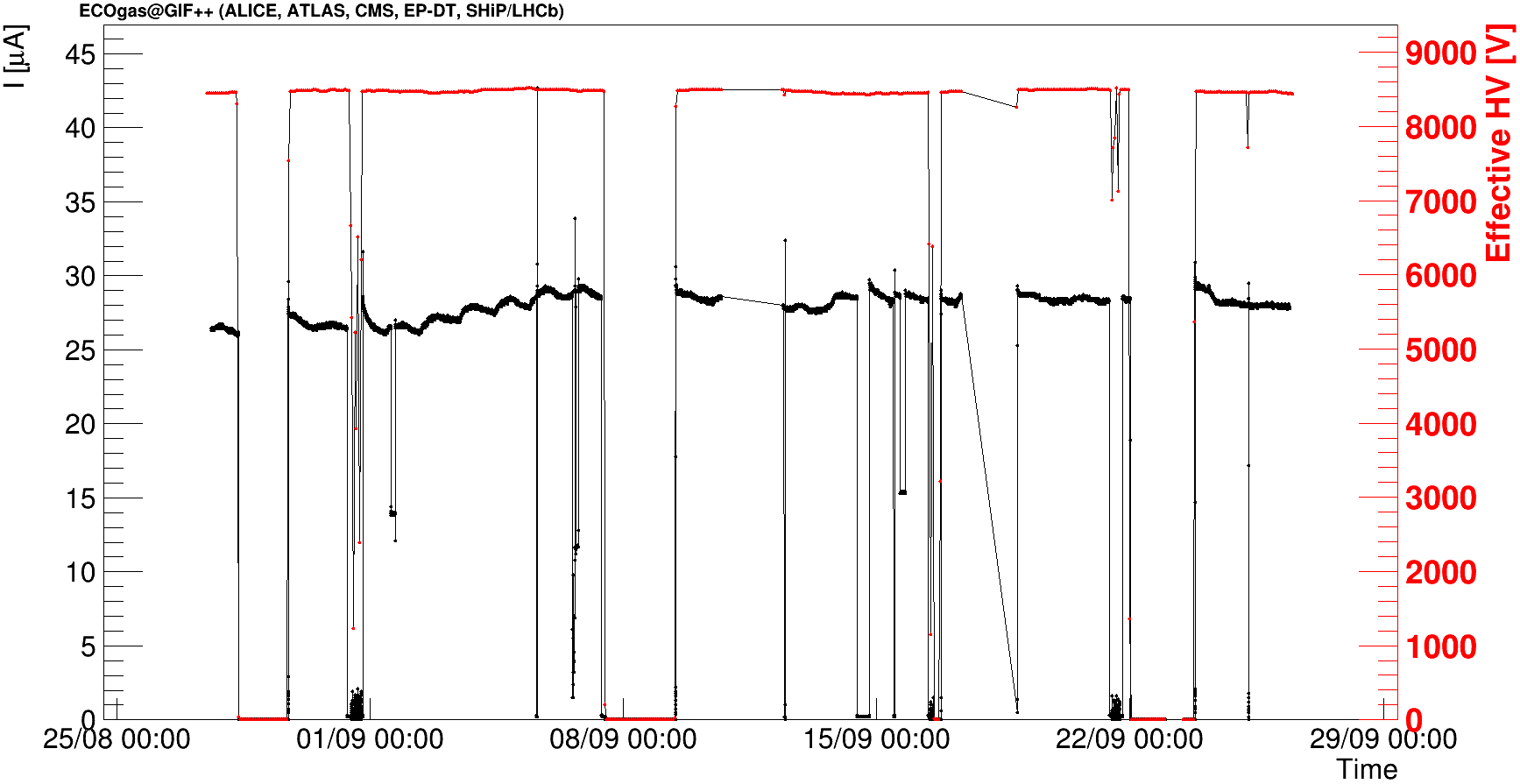}
	\caption{Trend of absorbed current and applied voltage of the SHiP RPC during aging test}
	\label{fig.stab}
\end{figure}

\begin{figure} [h!]
	\centering	
	\includegraphics[width = 1\linewidth]{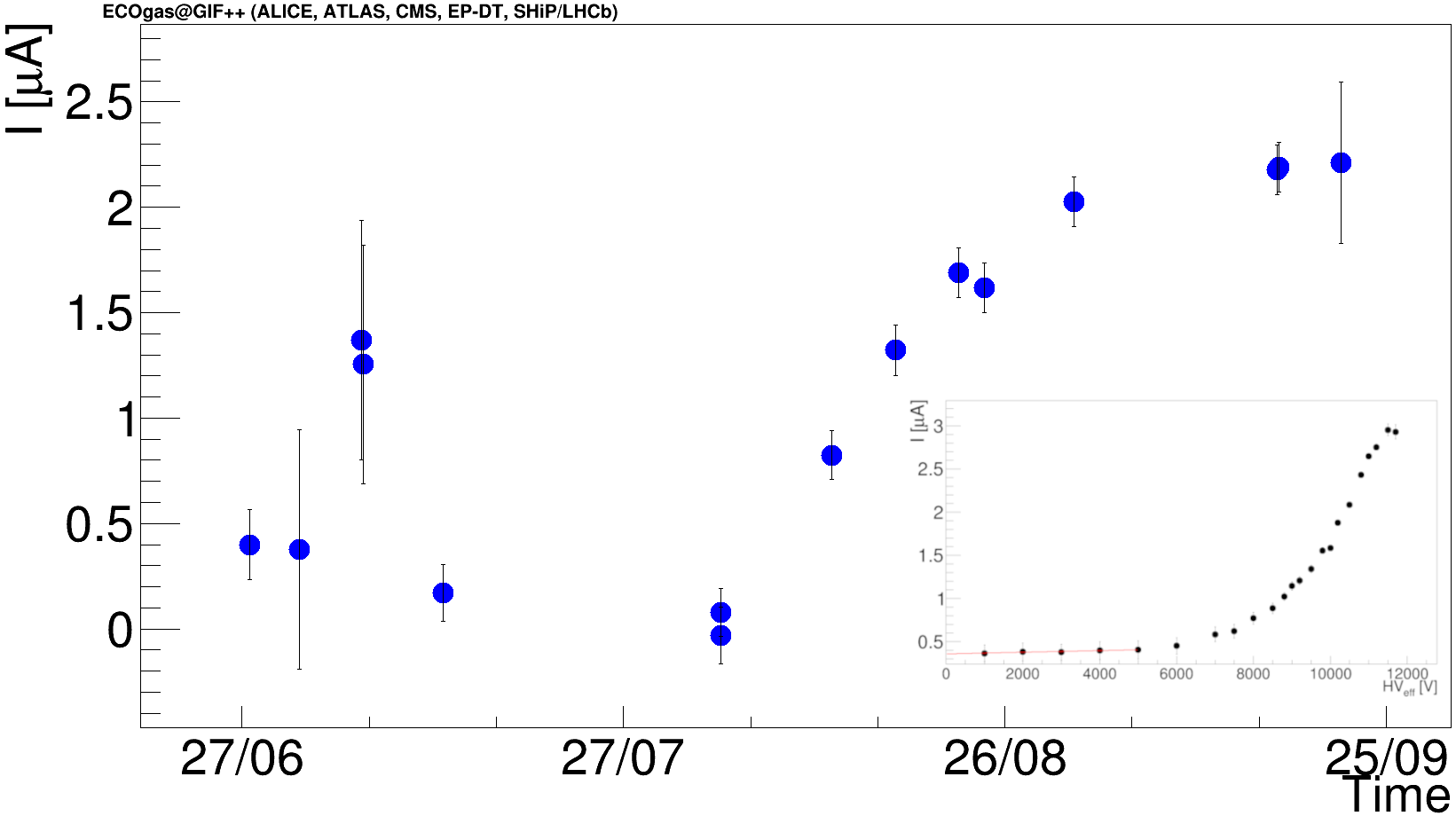}
	\caption{Results from CMS chamber. Big plot: current without Ohmic component in time, small plot: example of I(HV) scan. Currents are reported for source OFF}
	\label{fig.dark}
\end{figure}

At the GIF++, the nominal absorption factor is 2.2, corresponding to the highest rate achieved during the beam test (roughly 500 and 300 Hz/cm$^{2}$ for the RPCs closer and farther from the source). As shown in section \ref{sub.eff}, the currents absorbed with the eco-friendly gas mixtures are much higher with respect to the standard gas mixture. For this reason, it was decided to start the preliminary aging campaign with a value of voltage that corresponds to roughly 50\% efficiency. In this way, the absorbed current is lower than the one at the working point and, if any instability appears already at this lower efficiency, the irradiation can be stopped and the damages to the detectors will be reduced. The results shown in figure \ref{fig.stab} refer to the SHiP detector (1.6 mm single gas gap). For this RPC, the applied high voltage, corresponding to 50\% efficiency, is 8.7 kV, represented in the figure by the red line. The black line is, instead, the trend of the absorbed current. The applied high voltage is corrected for temperature and pressure\cite{tempCorr}, keeping the effective high voltage constant. The SHiP RPC shows a stable behavior after the integration of 18 mC/cm$^{2}$.
\vspace{-5pt}
\vspace{-5pt}
Once a week the gamma source is fully shielded (condition referred to as source off earlier) and one can measure the dark current (absorbed when there is no irradiation) of the detector and also monitor its stability over time. A drift of this quantity could be a sign of detector aging. Once a week we perform a high voltage scan and measure the dark current as a function of the applied voltage. An example of one of these scans can be seen in the small graph inside figure \ref{fig.dark}.

	
\vspace{-5pt}
\vspace{-10pt}
One can observe a first portion, where the current has a linear trend with the high voltage (Ohmic part) while, at around 8 kV, the trend becomes exponential, due to charge multiplication (physics current with Ohmic part). For this reason, the current in the multiplication region, is "contaminated" by a current that does not flow in the gas and, in order to get rid of it, we perform a linear fit in the range 0-4 kV and then we extrapolate the value of the Ohmic current at working point, and we subtract it to the measured current value to obtain the pure physics current. This value is shown, as a function of time, in the bigger graph inside figure \ref{fig.dark}.  The working point for the 2 mm RPCs has been set to 10.6 kV. Differently from the SHiP detector, the CMS RPC shows hints of an increasing trend in the absorbed current after integrating 10 mC/cm$^{2}$. This trend will be closely monitored in the future.
\vspace{-10pt}
\section{Conclusions}
\label{sec.conc}
\vspace{-10pt}
The RPC standard gas mixture contains F-gases, with a high global warming potential. New European Union regulations have imposed a progressive phase down in the production and usage of such gases. For this reason, the search for a new, more eco-friendly, gas mixture is an ongoing effort. The RPC EcoGas@GIF++ collaboration was born with the aim of testing the stability and performance of RPCs operated with said eco-friendly mixtures. In this framework, the result of some beam test studies have been discussed: two mixtures have been tested, ECO2 and ECO3 (with 60/35 and 69/25\% of \ce{CO_{2}} respectively). The performance of the detectors are promising, in terms of efficiency and cluster size. The current absorbed by the RPCs is twice as high as the one with the standard gas mixture. This parameter has to be closely monitored, to assess whether it can pose any threat for the long-term operation of the detectors. For this reason a systematic aging campaign is ongoing at GIF++.
\vspace{-10pt}
\nocite{*}
\bibliographystyle{elsarticle-num}
\bibliography{biblio.bib}
\vspace{-5pt}
\end{document}